%% file: ADMMmain.tex
\newcommand{\CLASSINPUTtoptextmargin}{0.73in}
\newcommand{\CLASSINPUTbottomtextmargin}{0.73in}
\newcommand{\CLASSINPUTinnersidemargin}{0.54in}

\documentclass[journal]{IEEEtran}
\usepackage{latexsym}

\usepackage{cite}

\ifCLASSINFOpdf
 \usepackage[pdftex]{graphicx}
\else
 \fi

\usepackage[cmex10]{amsmath}

\usepackage{algorithm}
\usepackage{algpseudocode}
\usepackage{array}
\usepackage{bm}   %added by Junyao
\usepackage{multirow} %added by Junyao
\usepackage{bigdelim}
 
\usepackage{booktabs}

\ifCLASSOPTIONcompsoc
  \usepackage[caption=false,font=normalsize,labelfont=sf,textfont=sf]{subfig}
\else
  \usepackage[caption=false,font=footnotesize]{subfig}
\fi

\usepackage{stfloats}

\ifCLASSOPTIONcaptionsoff
  \usepackage[nomarkers]{endfloat}
 \let\MYoriglatexcaption\caption
 \renewcommand{\caption}[2][\relax]{\MYoriglatexcaption[#2]{#2}}
\fi

\begin{document}
%
% paper title
% can use linebreaks \\ within to get better formatting as desired
% Do not put math or special symbols in the title.
\title{Enabling Distributed Optimization in Large-Scale Power Systems}
%
%
% author names and IEEE memberships
% note positions of commas and nonbreaking spaces ( ~ ) LaTeX will not break
% a structure at a ~ so this keeps an author's name from being broken across
% two lines.
% use \thanks{} to gain access to the first footnote area
% a separate \thanks must be used for each paragraph as LaTeX2e's \thanks
% was not built to handle multiple paragraphs
%

\author{Junyao~Guo,
        Gabriela~Hug,
        and~Ozan~Tonguz \vspace{-1.0cm}% <-this % stops a space

%\author{Junyao Guo, \IEEEmembership{Student Member, IEEE}, Gabriela Hug, \IEEEmembership{Senior Member, IEEE}, \\\ and Ozan K.~Tonguz, \IEEEmembership{Senior Member, IEEE}, \vspace{-0.8cm}
\thanks{J. Guo and O. K. Tonguz are with the Department
of Electrical and Computer Engineering, Carnegie Mellon University, Pittsburgh,
PA, 15213 USA (e-mail: junyaog@andrew.cmu.edu; tonguz@ece.cmu.edu)}
\thanks{G. Hug is with the Power Systems Laboratory, ETH Z{\"u}rich, Switzerland (e-mail: ghug@ethz.ch) }% <-this % stops a space
}

% make the title area
\maketitle

% As a general rule, do not put math, special symbols or citations
% in the abstract or keywords.

\begin{abstract}

Distributed optimization for solving non-convex Optimal Power Flow (OPF) problems in power systems has attracted tremendous attention in the last decade. Most studies are based on the geographical decomposition of IEEE test systems for verifying the feasibility of the proposed approaches. However, it is not clear if one can extrapolate from these studies that those approaches can be applied to very large-scale real-world systems. In this paper, we show, for the first time, that distributed optimization can be effectively applied to a large-scale real \textit{transmission network}, namely, the Polish 2383-bus system for which no pre-defined partitions exist, by using a recently developed partitioning technique. More specifically, the problem solved is the AC OPF problem with geographical decomposition of the network using the Alternating Direction Method of Multipliers (ADMM) method in conjunction with the partitioning technique. Through extensive experimental results and analytical studies, we show that with the presented partitioning technique the convergence performance of ADMM can be improved substantially, which enables the application of distributed approaches on very large-scale systems.

%Although distributed optimization has been widely studied in the context of power systems operation, little has been reported in the literature on its applicability to real-world non-convex power systems problems. To bridge the gap between theoretical studies and practice, this paper presents a partitioning technique derived from the Jacobian matrix of the Optimal Power Flow problem and Spectral Clustering which enables the application of distributed optimization to large-scale systems. The key contributions of this paper are the application of this method to a large-scale system, namely the Polish 2383-bus system and the evaluation of its performance with respect to convergence time and solution quality using the Alternating Direction Method of Multipliers (ADMM). Simulation results show that by choosing a good partition for the system, the convergence of the distributed method can be significantly improved, which enables the application of distributed approaches on large-scale systems. 
\end{abstract}
% Note that keywords are not normally used for peerreview papers.

% For peer review papers, you can put extra information on the cover
% page as needed:
% \ifCLASSOPTIONpeerreview
% \begin{center} \bfseries EDICS Category: 3-BBND \end{center}
% \fi
%
% For peerreview papers, this IEEEtran command inserts a page break and
% creates the second title. It will be ignored for other modes.
\IEEEpeerreviewmaketitle
{\renewcommand\baselinestretch{0.97}\selectfont
\vspace{-0.2cm}
\input{PolishIntro}
\vspace{-0.2cm}
\input{ACOPF}

\vspace{-0.2cm}
\input{ADMMAlgorithm}

\vspace{-0.2cm}
\input{PartitionMethod}
\vspace{-0.2cm}
\input{ADMMResult}

\vspace{-0.2cm}
\input{ADMMConclusion}

\vspace{-0cm}
\appendix[Optimality Conditions of ADMM Subproblems]
\label{App:A}
For the x-update in ADMM at the $\nu$-th iteration, the $k$-th region solves the following subproblem
\begin{subequations}
\label{subproblemk}
\begin{align}
\underset{{x}_{k}}{\text{minimize}}~~ & f_{k}({x}_{k})+{\lambda}_{k}^{\nu-1 \top}(A_{k}{x}_{k}-{z}_{k}^{\nu-1})\\
\nonumber &+\frac{1}{2}\|A_{k}{x}_{k}-{z}_{k}^{\nu-1}\|^{2}_{{\rho}_k^{\nu-1}}\\
\label{subgk}
\text{subject to} ~~& g(x_{k})=0.
\end{align}
\end{subequations}
For the simplicity of presentation, only equality constraints are considered in the following analysis which can be readily extended to include inequality constraints. The optimality conditions of subproblem (\ref{subproblemk}) are
\begin{subequations}
\begin{align}
\label{subLag}
&\nabla f_{k}({x}_{k})+\mu_{k}^{\top}\nabla g(x_{k})+A_{k}^{\top}{\lambda}_{k}^{\nu-1}\\
\nonumber&+A_{k}^{\top}\text{diag}({\rho}_k^{\nu-1})(A_{k}{x}_{k}-{z}_{k}^{\nu-1})=0\\
\label{subg}
&g(x_{k})=0,
\end{align}
\end{subequations}
where $\mu_{k}$ are the Lagrange multipliers of (\ref{subgk}). In (\ref{subLag}), the terms $A_{k}^{\top}{\lambda}_{k}^{\nu-1}$, $A_{k}^{\top}\text{diag}({\rho}_k^{\nu-1})A_{k}{x}_{k}$ and $A_{k}^{\top}\text{diag}({\rho}_k^{\nu-1}){z}_{k}^{\nu-1}$ only contain non-zero entries in the rows that correspond to the variables \mbox{$(x_{k}, \mu_{k})$} associated with the boundary buses. Hence, for the non-boundary buses, (\ref{subLag}) reduces to \mbox{$\nabla f_{k}({x}_{k})+\mu_{k}\nabla g(x_{k})=0$}, which combined with (\ref{subg}) are the centralized optimality conditions to be fulfilled at those non-boundary buses.

\vspace{-0.2cm}
\section*{Acknowledgment}
The authors would like to thank ABB for the financial support and particularly Dr. Xiaoming Feng for his invaluable comments. The authors would also like to thank Dr. Tomas Tinoco De Rubira for providing invaluable inputs  and Dr. Tomaso Erseghe for sharing details on his research on ADMM.

\vspace{-0cm}
\bibliographystyle{IEEEtran}
\bibliography{Partition}
\par}
\end{document}

%% file: PolishIntro.tex
\section{Introduction}
\label{ADMMintro}
In recent years, distributed optimization has received great attention for solving problems that arise in power systems operations, as it provides a promising alternative for solving complex optimization problems associated with grids that have a large number of distributed generation units \cite{6934985}. This technique allows dividing an optimization problem into subproblems associated with different regions of the grid, which are solved separately and simultaneously with periodic information exchanges.

One key application considered for distributed optimization is the Optimal Power Flow (OPF) problem, which is at the heart of power systems operations and planning. OPF problems have been studied for over half a century \cite{cain2012history}. The most common objective of OPF is to schedule the generation units optimally subject to the power flow balances and operational constraints such as transmission line capacities. Distributed methods based on various decomposition techniques have been proposed to solve the OPF problem, including Augmented Lagrangian Relaxation \cite{baldick1999fast}\cite{kim2000comparison}, or more specifically ADMM \cite{erseghe2015distributed} and Optimality Condition Decomposition (OCD) \cite{conejo2002decomposition}, which have been shown to have fast convergence rates and achieve a solution that equals or is close to the local optimum. Even though such results seem promising, the experiments in these studies are conducted either on IEEE standard test systems \cite{erseghe2015distributed}\cite{conejo2002decomposition} whose size and complexity are not comparable with those of real power systems, using several connected small-scale utility networks \cite{kim2000comparison} where a decomposition of the system can be trivially identified or using a given partition \cite{baldick1999fast} which may lead to a suboptimal performance of the distributed algorithm. Hence, these studies do not address the important question of how to partition a general large-scale network in a way that is suitable for distributed optimization. 
%This leaves the following question open: if given a large-scale network with no obvious separable regions, how can one apply distributed methods and whether distributed methods can still achieve good convergence performances? 
In fact, due to the non-convexity introduced by the AC power flow equations, OPF is known to be a difficult problem to solve even in a centralized manner for large systems \cite{lam2012distributed}. Further difficulties arise in the search for a distributed solution since optimality or even convergence cannot be guaranteed for most distributed methods on non-convex problems \cite{conejo2006decomposition}\cite{boyd2011distributed}. 

While the IEEE standard test systems serve as good platforms to showcase the functionality of new algorithms, researchers still face a practical yet challenging question: can one successfully apply distributed algorithms to real-world large-scale power systems problems, in particular to non-convex OPF problems? Ultimately, the value of distributed optimization can only be understood and its full potential can only be realized if it can be applied to large-scale real power systems.
%Other distributed methods proposed in the literature for solving non-convex optimization problems include those described in \cite{rahbari2014cooperative} and \cite{binetti2014distributed}. However, these methods rely on approximating non-convex functions and are only suitable for specific problem formulations. 

%In our previous work, we have developed a distributed optimization framework that consists of an intelligent partitioning method to define the power system partition and a decomposition method to solve the optimization problem in a distributed fashion. This framework has been tested on multiple IEEE standard test systems to solve the non-convex AC OPF problem and multi-step AC OPF problem with the integration of renewable energy and storage. Simulation results on the test systems show that distributed optimization can achieve improved efficiency compared with the centralized approach if the partition of the system is devised intelligently and direct communications between operational regions is supported. 

Several practical issues arise when implementing distributed methods on large-scale networks. One critical issue is the system partitioning, which has been shown to have considerable impact on the performance of the distributed algorithm \cite{junyao2015impact}. In \cite{guointelligent}, we proposed a partitioning method based on spectral clustering that significantly improves the convergence speed of OCD using IEEE benchmark systems. This partitioning method defines an affinity metric that captures the coupling among the regions and maps the power system partitioning problem to a graph partitioning problem. Yet, it remains to be seen whether this partitioning technique can devise good partitions of real-world systems for an efficient implementation of different distributed methods. Hence, in this paper, we attempt to apply this partitioning technique in conjunction with the ADMM method to a non-convex OPF problem in the Polish 2383-bus system. Specifically, we first use the partitioning method to define subregions of the network, and then apply ADMM, which is a state-of-the-art distributed method that has been shown to have fast convergence for solving the OPF problem \cite{erseghe2015distributed}. The main contributions of this paper are the following:
%Using ADMM with the proposed partitioning method allows evaluating the usefulness of this method for other distributed algorithms than the one for which it was originally designed, namely OCD. 
\begin{itemize}
\item To show that a recently developed partitioning technique enables the very efficient solution of non-convex OPF problem in a large-scale real transmission network using a distributed optimization approach as opposed to a centralized approach. This result is a milestone as it shows for the first time that distributed optimization is a practically viable approach to key optimization problems in large-scale real power systems with no pre-defined partitions.
\item To show that the promising partitioning technique that was recently developed is applicable to other distributed methods such as ADMM (as opposed to OCD) as well. This provides initial evidence that the developed partitioning technique is sufficiently general to work with various distributed optimization algorithms.
%We carry out the full process of partitioning the grid to implementing a distributed optimization approach on the Polish system and show experimentally that a near-optimal point can be found for the OPF problem in a reasonable amount of time. 
\end{itemize}

The rest of the paper is organized as follows. Section \ref{relatedwork} summarizes related work. While Section \ref{ACOPF} formulates the general AC OPF problem, Section \ref{ADMMalgorithm} describes the distributed OPF method based on ADMM. Section \ref{Partition} describes the network partitioning technique proposed in a previous work and why it can work effectively with ADMM. Section \ref{ADMMresult} presents extensive simulation results obtained from applying the distributed approach to the Polish system, and Section \ref{discussion} provides physical interpretations of the main results and future directions. Finally, Section \ref{ADMMconclusion} concludes the paper.
\vspace{-0.2cm}
%as the merit and the full potential of distributed optimization can only be realized and appreciated if it can be put into practical use in real-world systems.
\section{Related Work}
\label{relatedwork}
To overcome some of the difficulties in solving non-convex OPF in a distributed manner, one approach proposes convexifying the OPF problem before applying a distributed algorithm \cite{lam2012distributed}\cite{erseghe2013power, dall2013distributed, magnusson2014distributed}. A common approach for convexification is based on semidefinite relaxation \cite{lavaei2012zero}\cite{madani2015convex}. However, due to lack of adherence to the original non-convex problem, the optimality achieved by this approach can only be ensured for some simple networks such as the IEEE benchmark systems and acyclic radial networks \cite{lavaei2012zero}\cite{madani2015convex}. Moreover, semidefinite programming solvers generally result in large computational efforts, hence, they do not scale well to problems in large transmission networks, which typically have thousands of buses and meshed topologies \cite{erseghe2014distributed}. 

In addition, there are several other studies that demonstrate the scalability of the proposed distributed methods. In \cite{phan2014some}, two decomposition algorithms for solving the Security-Constrained OPF problem are investigated and applied to the Polish 3012-bus system. However, the OPF problem is decomposed according to contingency scenarios and not geographical regions, which is much harder. For optimizing distribution networks, the authors in \cite{peng2014distributed} propose using second-order cone relaxation for convexifying the OPF problem before applying a distributed algorithm, and test this approach on a real-world 2065-bus distribution system. As transmission networks are meshed networks, it is not clear whether the results obtained in \cite{peng2014distributed} still hold for these networks. In \cite{kraning2014dynamic}, a method based on ADMM and proximal message passing is proposed for solving a dynamic OPF problem, and is tested on networks with thousands of nodes. However, both the nodes and the branches are generated randomly, which could result in a network that is very different from real-world systems. Unlike the aforementioned studies, this paper contributes to the geographical decomposition of a large-scale real transmission network for implementing distributed methods. 

In terms of power system partitioning, other methods are proposed in \cite{sanchez2014hierarchical}\cite{cotilla2013multi} that are based on hierarchical clustering and electrical distances, which, however, are not specifically designed for distributed optimization. Hence, the spectral partitioning approach used in this paper is a better choice for the considered problem \cite{guointelligent}. 

%% file: ACOPF.tex
\section{AC OPF Problem}
\label{ACOPF}

The standard AC OPF problem is considered in this paper, where the objective is to minimize the generation cost and the constraints are the power flow balances, the generation capacities and the limits on the voltage magnitudes. Mathematically, the problem is given by:
\begin{subequations}
\label{centOPF}
\begin{align}
\label{eq1}
\underset{V, P, Q} {\text{minimize}}~~&f({P})=\sum_{i=1}^{n_{b}} \left(a_{i}P_{i}^2+b_{i}P_{i}+c_{i}\right) \\
%\end{equation}
%subject to
%\begin{align}
\label{pf}
\text{subject to}~~&P_{i}+jQ_{i}-P_{i}^{d}-jQ_{i}^{d}=V_i\sum\limits_{j\in {{\Omega }_{i}}}Y_{ij}^*V_j^*\\
\label{eqPlimit}
&P_{i}^{\min}\leq P_{i}\leq P_{i}^{\max}\\
\label{eqQlimit}
&Q_{i}^{\min}\leq Q_{i}\leq Q_{i}^{\max}\\
\label{eqVlimit}
&V_{i}^{\min}\leq |V_{i}|\leq V_{i}^{\max},
%|I_{ij}|^{2}\leq &|I_{ij,max}|^{2}
\end{align}
\end{subequations}
for $i=1, \ldots, n_{b}$. Here, $n_{b}$ is the number of buses, and $(a_i,b_i,c_i)$ are the cost parameters of generator at bus $i$. $(V_{i}, P_{i}, Q_{i})$ are the complex voltage, the active power output and the reactive power output of generator at bus $i$. $(P_{i}^{d},Q_{i}^{d})$ are the active and reactive load at bus $i$, $Y_{ij}$ is the $ij$-th entry of the nodal admittance matrix, and $\Omega_i$ is the set of buses connected to bus $i$.
%The notation used in the problem formulation is described below:
%
%\begin{tabular}[h]{ll}
%$n_{b}$ & number of buses\\
%$a_i,b_i,c_i$ & cost parameters of generator $i$ \\
%$P_{i}, Q_{i}$ & active and reactive power output of generator\\
%& at bus $i$ \\
%$P_{i}^{d},Q_{i}^{d}$&active and reactive load at bus $i$\\
%$V_{i}$&complex voltage at bus $i$\\
%$Y_{ij}$&$ij$ entry of nodal admittance matrix\\
%$\Omega_i$ & set of buses connected to bus $i$ 
%%$I_{ij}$&current on line from bus $i$ to bus $j$\\
%\end{tabular}
Apart from constraints (\ref{pf})-(\ref{eqVlimit}), the angle of the reference bus is set to zero. We note that line thermal limits have been omitted here just to keep the presentation simple, but can be readily added to the problem formulation. In fact, in Section \ref{ResultTL}, line limits are added in the numerical experiments and their impact on the convergence of the distributed method is evaluated.

%% file: ADMMAlgorithm.tex
\section{ADMM-Based Distributed OPF}
\label{ADMMalgorithm}

\subsection{OPF Formulation for Distributed ADMM}
\label{distributedADMM}
To decompose the OPF problem (\ref{centOPF}), the power system is first partitioned into smaller regions, and a local OPF problem is formulated for each region. Then, to solve the decomposed OPF problem in a distributed way, an ADMM method with refined iterations is used \cite{erseghe2015distributed}. For the following analysis, we use $K$ to denote the total number of regions and $\mathcal{R}_{k}, k=1,..., K,$ to denote the set of buses assigned to region $k$ with $\mathcal{R}_{k}\cap\mathcal{R}_{l}=\emptyset, \forall l\neq k$. The set $\mathcal{V}_{k}$ is also introduced to denote the joint set of buses including the buses in $\mathcal{R}_{k}$ and the buses in neighboring regions that are directly connected to buses in $\mathcal{R}_{k}$.

\begin{figure}[t]
\setlength{\abovecaptionskip}{0cm} 
\centering
\includegraphics[trim = 0mm 0mm 0mm 0mm, clip=true,width=7cm]{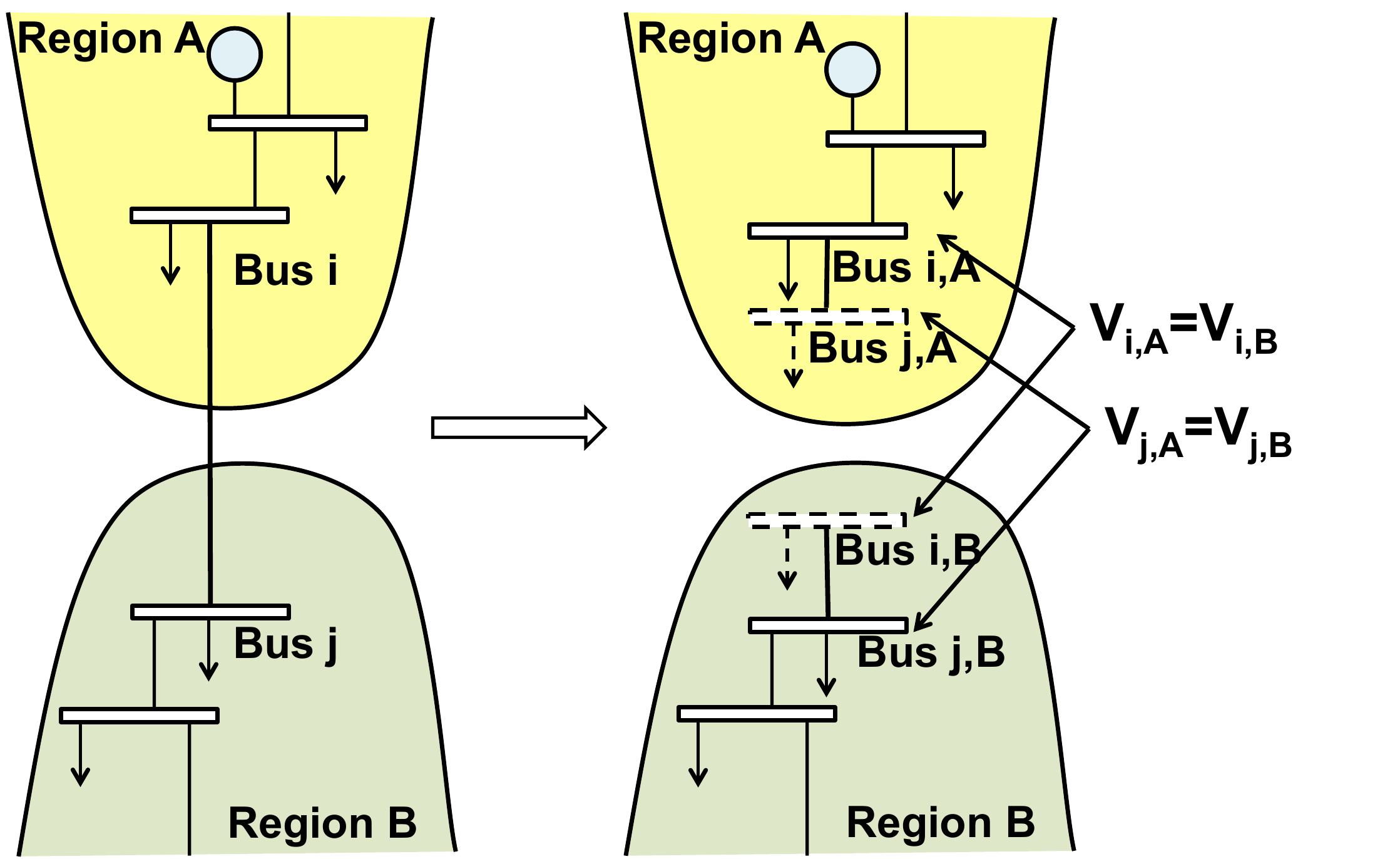} 
\caption{Duplicating voltages at boundaries of regions.}
\label{fig:buscopy}
\end{figure}

To enable the distributed approach, the power system is decoupled by duplicating the voltages at the boundary buses of each region. Constraints are then added to enforce the duplicate voltages to be equal to one another. Figure \ref{fig:buscopy} illustrates this decoupling of the system where bus $i$ and bus $j$ are the buses that are connected by tie line $ij$. The voltages at bus $i$ and bus $j$ are duplicated, and the copies assigned to Region A are denoted by $V_{i,A}$ and $V_{j,A}$. Similarly, Region B is assigned the copies $V_{i,B}$ and $V_{j,B}$. To ensure equivalence with the original problem, the constraints $V_{i,A}=V_{i,B}$ and $V_{j,A}=V_{j,B}$ are added to the problem. These operations remove the tie line and separate the two regions. As discussed in \cite{erseghe2015distributed}, the added constraints are equivalent to
\vspace{-0.15cm}
\begin{equation}
\begin{aligned}
V_{i,A}-V_{j,A}&=V_{i,B}-V_{j,B}\\
V_{i,A}+V_{j,A}&=V_{i,B}+V_{j,B}.
\end{aligned}
\label{xAB}
\end{equation}
\vspace{-0.15cm}

For each tie line $ij$ with $i \in \mathcal{R}_{A}, j \in \mathcal{V}_{A}\setminus \mathcal{R}_{A}$ we further introduce two auxiliary variables $z_{i,j}^{+}$ and $z_{i,j}^{-}$ to Region A, and two auxiliary constraints
\vspace{-0.15cm}
\begin{equation}
z_{i,j}^{-}=\beta^{-}(V_{i,A}-V_{j,A}),~~~~~z_{i,j}^{+}=\beta^{+}(V_{i,A}+V_{j,A}),
\label{eq:zV}
\end{equation}
where $\beta^{-}$ and $\beta^{+}$ are scaling factors. Constant $\beta^{-}$ is set to be larger than $\beta^{+}$ to give more weight to $V_{i,A}-V_{j,A}$, which is strongly related to the line flow through tie line $ij$ \cite{erseghe2015distributed}. Similarly, two auxiliary variables $z_{j,i}^{-}$ and $z_{j,i}^{+}$ are introduced to Region B. Hence, the feasible region of all the $z$'s associated with tie lines is defined as
\begin{equation}
\mathcal{Z}=\{({z^{-}},{z^{+}})~|~ z_{i,j}^{-} =-z_{j,i}^{-}, ~z_{i,j}^{+} =z_{j,i}^{+}, \forall (i,j) \in \mathcal{B}\},
\label{zfeasible}
\end{equation}
where $\mathcal{B}$ is the set of inter-region tie lines. Set (\ref{zfeasible}) is derived from expressing the constraints in (\ref{xAB}) using the definitions given in (\ref{eq:zV}).  Let  \mbox{${x}_{k}=\{(V_i, P_{i}, Q_{i})~|~i \in \mathcal{V}_{k}\}$} and \mbox{${z}_{k}=\{ (z_{i,j}^{-},  z_{i,j}^{+})~|~i \in \mathcal{R}_{k}, j \in \mathcal{V}_{k}\setminus \mathcal{R}_{k}\}$} denote all the primal variables and auxiliary variables associated with the buses in region $k$, respectively. The OPF problem can then be expressed in terms of variables assigned to different regions as follows:
\vspace{-0.2cm}
\begin{subequations}
\label{eq:OPF}
\begin{align}
\label{diseq1}
\underset{{x}, {z}} {\text{minimize}} ~~~&\sum_{k} f_{k}({x}_{k})\\
\label{diseq2}
\text{subject to}~~~& A_{k}{x}_{k}={z}_{k},~\forall k\\
\label{diseq3}
&{x}_{k} \in \mathcal{X}_{k}, ~\forall k\\
\label{diseq4}
& {z} \in \mathcal{Z},
\end{align}
\end{subequations}
where $f_{k}({x}_{k})$ is the generation cost in region $k$ and constraint (\ref{diseq2}) is obtained by expressing (\ref{eq:zV}) using $x_{k}$ and $z_{k}$. Constraint (\ref{diseq3}) enforces the local feasibility constraints, namely, constraints (\ref{pf})-(\ref{eqQlimit}) for $\forall i \in\mathcal{R}_{k}$ and constraint (\ref{eqVlimit}) for $\forall i \in\mathcal{V}_{k}$.
%\begin{equation}
%\begin{aligned}
%P_{G_{i}}+jQ_{G_{i}}-&P_{D_{i}}-jQ_{D_{i}}=V_i\sum\limits_{k\in {{\Omega }_{i}}}Y_{ij}^*V_j^*, ~\forall i \in\mathcal{R}_{k} \\
%&V_{i}^{\min}\leq |V_{i}|\leq V_{i}^{\max}, \forall i \in\mathcal{V}_{k}\\
%&P_{G_{i}}^{\min}\leq P_{G_{i}}\leq P_{G_{i}}^{\max}, \forall i \in\mathcal{R}_{k}\\
%&Q_{G_{i}}^{\min}\leq Q_{G_{i}}\leq Q_{G_{i}}^{\max}, \forall i \in\mathcal{R}_{k}.
%\end{aligned}
%\end{equation}
An important property of problem (\ref{eq:OPF}) is that if ${z}$ is fixed, then problem (\ref{eq:OPF}) can be decomposed into subproblems where each subproblem only contains the local variables ${x}_{k}$. This property enables distributing the computations of ADMM to solve problem (\ref{eq:OPF}), as described next. 

\subsection{Distributed ADMM Algorithm}
The ADMM algorithm minimizes the Augmented Lagrangian function of (\ref{eq:OPF}), which is given as follows \cite{erseghe2015distributed}:
\begin{equation}
\begin{aligned}
L({x},{z},{\lambda})=&\sum_{k} \Big\{ f_{k}({x}_{k})+{\lambda}_{k}^{\top}(A_{k}{x}_{k}-{z}_{k})\\&+\frac{1}{2}\|A_{k}{x}_{k}-{z}_{k}\|^{2}_{{\rho}_k} \Big\},
\end{aligned}
\label{eq:Aug}
\end{equation}
where $\|{x}\|^{2}_{{\rho}}={x}^{\top} \text{diag} ({\rho}) {x}$ is the square of a weighted norm of ${x}$. The vector ${\rho}$ is a vector of penalty parameters whose entries are increased during the iterative process to ensure convergence of ADMM\cite{erseghe2015distributed}. With (\ref{eq:Aug}) formulated, the $\nu+1$-th iteration of ADMM consists of the following steps:
\begin{subequations}
\label{eq:ADMMIter}
\begin{align}
\label{xupdate}
{x}^{\nu+1} &= \text{argmin}_{{x} \in \mathcal{X}}~~ L({x},{z}^{\nu},{\lambda}^{\nu})\\
{z}^{\nu+1}&=\text{argmin}_{{z}\in \mathcal{Z}}~~L({x}^{\nu+1},{z},{\lambda}^{\nu})\\
{\lambda}^{\nu+1}&={\lambda}^{\nu}+\text{diag}({\rho}^{\nu})(A{x}^{\nu+1}-{z}^{\nu+1}).
\end{align}
\end{subequations}
The ${z}$ update involves information exchanges among regions, and can also be computed locally once the updated information from neighboring regions is acquired. Hence, the ADMM iterations can be carried out in a completely distributed fashion with only local information exchanges but no centralized coordination. Updating ${x}$ requires solving non-convex subproblems, while updating ${z}$ solves a quadratic programming problem and updating ${\lambda}$ is trivial. 

To enhance the performance of ADMM on non-convex problems, the penalty parameter ${\rho}$ is usually updated to make the Augmented Lagrangian function convex near the solution. Specifically, for any region $k$, ${\rho}_{k}$ is updated as follows \cite{erseghe2015distributed}: 
\begin{subequations}
\begin{align}
\label{eq:rhocompute}
\tilde{{\rho}}^{\nu+1}_k=& \left\{ \begin{array}{cl}
\|{\rho}^{\nu}_k\|_{\infty} \bm{1} & \text{if~~} \Gamma_k^{\nu+1}\leq \gamma \Gamma_k^{\nu}\\
\tau\|{\rho}^{\nu}_k\|_{\infty} \bm{1}  & \text{otherwise}
\end{array}\right. \\
\label{eq:rhoupdate}
\rho_{k,i,j}^{\nu+1}=&\max\{\tilde{\rho}_{k,i,j}^{\nu+1},\tilde{\rho}_{l,j,i}^{\nu+1}\} \end{align}
\end{subequations}
with constants $0<\gamma<1$ and $\tau>1$, and with $\bm{1}$ denoting the all-ones vector. Equation (\ref{eq:rhoupdate}) holds for $\forall l \neq k$ and $(i,j) \in \mathcal{B}$ such that $i \in \mathcal{R}_{k}$ and $j \in \mathcal{R}_{l}$. \mbox{$\Gamma_k^{\nu+1}=\|A_{k}{x}_{k}^{\nu+1}-{z}_{k}^{\nu+1}\|_{\infty}$} is defined as the local primal residue\cite{boyd2011distributed}, which measures the error in the coupling constraints. Parameter ${\rho}$ is updated after the ${\lambda}$-update in (\ref{eq:ADMMIter}). First, an individual ${\rho}_{k}$ is updated for each region via (\ref{eq:rhocompute}). Then the $\rho$ associated with each tie line is adjusted via (\ref{eq:rhoupdate}). 
%Notice that (\ref{eq:rhoupdate}) requires the exchange of $\tilde{\rho}_{k,i,j}$ and $\tilde{\rho}_{l,j,i}$ between regions $k$ and $l$ that the tie line $ij$ connects. To reduce the communication cost, this exchange can be implemented during the information exchange before the ${z}$-update. This may lead to a slightly degraded performance of ADMM since the penalty parameters are not updated immediately, which, on the other hand, only requires one information exchange for each ADMM iteration.

A detailed procedure of the distributed ADMM algorithm is illustrated in Algorithm \ref{algADMM}, where $\nu$ denotes the iteration counter. A general way to check the convergence of (\ref{eq:ADMMIter}) is to check whether the primal residue ($\Gamma_k, \forall k$) is smaller than some $\epsilon$ \cite{boyd2011distributed}. However, in the AC OPF problem, power balance feasibility must also be ensured. This feasibility is checked after averaging the duplicate voltages in each iteration. Convergence is declared when both the primal residue and the maximum bus power mismatch after voltage averaging fall below $\epsilon$. The convergence of this ADMM approach is proved in \cite{erseghe2015distributed} with the assumption that a local minimum can be identified when solving the local OPF problems.

Some guidelines regarding the choice of parameters $\beta^{+}, \beta^{-}, \gamma, \tau$, and the initial value of $\rho$ are provided in \cite{erseghe2015distributed}.
%\begin{itemize}
%\item $\beta^{+}$ and $\beta^{-}$ should be chosen in a way that $\beta^{-}>\beta^{+}>0$ to emphasize more on the power flow on the tie lines. 
%\item For the update of penalty parameters ${\rho}$, $\gamma$ is usually set to a value slightly lower than 1, and $\tau$ is set to a value slightly larger than 1 to avoid rapid increase of the penalty parameter. The larger $\tau$ is, the faster ${\rho}$ increases and consequently the faster ADMM converges, which, however, generally leads to solutions with worse quality since the algorithm proceeds more aggressively. Hence, there is a trade-off between the convergence speed of ADMM and the solution quality it achieves. \item For the initial value of $\rho_{0}$, it can be set to a value of the same or slightly larger magnitude as the initial objective function value if a good starting point is used ; otherwise, it should be set to a much smaller value compared with the initial objective function to ensure convergence.
%\end{itemize}
We should mention that the exact parameters used in ADMM are usually tuned through empirical studies. The reader is referred to  \cite{ghadimi2015optimal}\cite{shi2014linear} for techniques on tuning ADMM parameters for specific problems.
\begin{algorithm}[t]
\caption{Distributed OPF in Region $k$}
\label{algADMM}
\begin{algorithmic}[1]
\State \textbf{Initialization} Initialize ${x}_{k}^{0}$, ${z}_{k}^{0}={0}$, ${\lambda}_{k}^{0}={0}$, ${\rho}_{k}^{0}=\rho_{0} \bm{1}$, $\nu=0$ 
\While{Not converged}
\State $\nu \leftarrow \nu+1$
\State Update ${x}_{k}$ by solving the local OPF
\begin{align}
\nonumber {x}_{k}^{\nu} = & \underset{{x}_{k}\in \mathcal{X}_{k}}{\text{argmin}} ~f_{k}({x}_{k})+{\lambda}_{k}^{\nu-1 \top}(A_{k}{x}_{k}-{z}_{k}^{\nu-1})\\
\nonumber &+\frac{1}{2}\|A_{k}{x}_{k}-{z}_{k}^{\nu-1}\|^{2}_{{\rho}_k^{\nu-1}}
\end{align}
\State Prepare messages ${m}_{k}^{\nu}=A_{k}{x}_{k}^{\nu}$
\State Broadcast ${m}_{k}^{\nu}$ to neighboring regions and receive ${m}_{l}^{\nu}$ \indent from each neighboring region $l \neq k$
\State Update ${z}_{k}$ using
\begin{align}
\nonumber z_{i,j}^{- \nu}&=\frac{1}{2}(m_{k,i,j}^{-\nu}-m_{l,j,i}^{-\nu})\\
\nonumber z_{i,j}^{+ \nu}&=\frac{1}{2}(m_{k,i,j}^{+\nu}+m_{l,j,i}^{+\nu})
\end{align}
\State Update ${\lambda}_{k}$ using
\begin{equation}
\nonumber{\lambda}_{k}^{\nu}={\lambda}_{k}^{\nu-1}+\text{diag}({\rho}^{\nu-1}_k)(A_{k}{x}_{k}^{\nu}-{z}_{k}^{\nu})
\end{equation}
\State Calculate the primal residue $\Gamma_k^{\nu}$ for each region $k$
\State Check convergence
\State Compute $\tilde{{\rho}}_{k}^{\nu}$ according to (\ref{eq:rhocompute})
%\begin{equation}
%\nonumber \tilde{{\rho}}^{\nu}_k= \left\{ \begin{array}{cl}
%\|{\rho}^{\nu-1}_k\| {1} & \text{if~~} \Gamma_k^{\nu}\leq \gamma \Gamma_k^{\nu-1}\\
%\tau\|{\rho}^{\nu-1}_k\| {1}  & \text{otherwise}
%\end{array}\right.
%\end{equation}
\State Broadcast $\tilde{{\rho}}_{k}^{\nu}$ to neighboring regions and receive $\tilde{{\rho}}_{l}^{\nu}$ \indent from each neighboring region $l \neq k$
\State Update ${\rho}_{k}$ using (\ref{eq:rhoupdate})
%\begin{equation}
%\nonumber\rho_{k,i,j}^{\nu}=\max\{\tilde{\rho}_{k,i,j}^{\nu},\tilde{\rho}_{l,j,i}^{\nu}\}
%\label{eq:rhoupdate}
%\end{equation}
\EndWhile
\end{algorithmic}
\end{algorithm}

%% file: PartitionMethod.tex
\section{Power System Partitioning}
\label{Partition}
The performance of distributed methods depends highly on how the problem is partitioned \cite{junyao2015impact}. Hence, to enable an effective application of ADMM, the partitioning method first proposed in \cite{guointelligent} is used, which is denoted as spectral partitioning in the remainder of this paper. Based on the premise that weaker coupling between regions results in improved performance of distributed approaches, the spectral partitioning method first defines an affinity metric that captures the coupling between buses. An affinity matrix is constructed from the pairwise affinities between buses, and spectral clustering \cite{ng2002spectral} is applied to this matrix for identifying groups of buses that are strongly coupled. 
%Please note, while the partitioning method has been presented in \cite{guointelligent}, it was derived from OCD and tested for this decomposition method on the traditional IEEE benchmark systems. Here, the intention is to demonstrate the general applicability and scalability of our approach.

The detailed procedure of spectral partitioning is illustrated in Algorithm \ref{algIP}. Here, $N$ denotes the number of trials of executing the K-means clustering method using different initial centroids of the clusters. As multiple partitions are found due to different initializations of K-means, the most balanced partition, i.e., the one with the smallest largest region is chosen to be the best partition of the system. The reason for this choice is that the subproblems need to be solved in parallel before each information exchange occurs, thus a balanced partition is preferable for reducing the maximum subproblem solution time in each ADMM iteration. 
%Note that this criterion for choosing the best partition is different from the spectral radius criterion used in \cite{guointelligent}, where the partitioning technique is applied to OCD whose convergence is dependent on the spectral radius.

Here, we describe one way to compute the affinity matrix, which is based on the evaluation of the first-order optimality conditions using the distributed ADMM approach. Let \mbox{$F(y)=0$} denote the first-order optimality conditions associated with the centralized problem formulated in (\ref{centOPF}) where $y$ includes both the primal and dual variables, $H$ denote the Jacobian of $F(y)$ evaluated at the centralized optimal point $y^{c}$, and $y^{d}$ denote the counterparts to the centralized solution obtained after the x-update (\ref{xupdate}) in ADMM at the $\nu$-th iteration. As $y^{d}$ is generally different from $y^{c}$, we would like to evaluate how well the distributed solution $y^{d}$ satisfies the centralized optimality conditions. An important observation here is that the optimality conditions associated with the non-boundary buses derived for the local OPF problems are identical to those associated with the non-boundary buses derived for the centralized OPF problem, which is shown explicitly in Appendix \ref{App:A}. Therefore, these optimality conditions can be satisfied under the assumption that a local minimum can be found for the local OPF problems. Furthermore, it can be easily checked that the primal feasibilities (\ref{eqPlimit})-(\ref{eqVlimit}) at all buses are satisfied by solving local OPF problems. Hence, it only remains to be evaluated how optimality conditions apart from (\ref{eqPlimit})-(\ref{eqVlimit}) associated with the boundary buses are satisfied by $y^{d}$. Denoting this part of optimality conditions by $F^{b}(y)=0$ and assuming that $\nu$ is large enough such that $y^{d}$ is in the vicinity of $y^{c}$, then \mbox{$F^{b}(y^{d})$} can be approximated by \mbox{$F^{b}(y^{d})\approx H^{b}(y^{d}-y^{c})$}. Here, $H^{b}$ is constructed by taking all the rows in $H$ that contain any non-zero entry $H_{m,n}$ where its associated variables $y_{m}$ and $y_{n}$ belong to different subproblems. It is expected that if the entries in $H^{b}$ are generally small in terms of their absolute values, then $y^{d}$ can better satisfy the centralized optimality conditions, which, on the other hand, shows that ADMM can converge easier to an optimal point with possibly fewer iterations. 

Hence, we aim to find a partition such that $H^{b}$ contains mostly small values. Since minimizing the entries in $H^{b}$ exactly is generally hard, a heuristic way is to consider $|H_{m,n}|$ as an affinity measure between variables $y_{m}$ and $y_{n}$ in spectral partitioning. If $|H_{m,n}|$ is large, $y_{m}$ and $y_{n}$ are more likely to be grouped into one subproblem, hence $H_{m,n}$ is less likely to appear in $H^{b}$ based on the construction of $H^{b}$.

Apart from using the Jacobian matrix $H$ of the optimality conditions, the affinity metric should also capture the electrical characteristics of the system, which can be represented by the admittance matrix $Y$. Hence, the affinity matrix $A$ used in this study is constructed by combining the Jacobian matrix $H$ and the admittance matrix $Y$. As there is usually more than one variable associated with any bus, the affinity between buses $i$ and $j$ is computed by 
\vspace{-0.1cm}
\begin{equation}
\vspace{-0.1cm}
A_{i,j}=\underset{m\in S_{i}}\sum\underset{n\in S_{j}}\sum|H_{m,n}|+|Y_{i,j}|,
\label{eq:affinity}
\end{equation}
where $S_{i}$ and $S_{j}$ denote the sets of the indices of the variables associated with buses $i$ and $j$, respectively.
%The admittance matrix represents the electrical structure of the systems, while the Hessian matrix contains additional information such as the Lagrange multipliers of the constraints, hence represents the computational coupling between the buses. 
%that even though $H$ needs to be evaluated at the optimal solution, it is only computed once for a specific operating point which requires limited empirical knowledge of the system operation. The entries in $H$ are not expected to vary considerably with the operating points as the entries are functions of the line admittance, the voltage magnitude, the \(sin\) and \(cos\) of the differences between two bus angles, and the Lagrange multipliers of power flow constraints which all change within small ranges. 
Note that although $H$ has to be evaluated at the optimal point for a specific operating point, it has been shown that the partition found by using the $H$ evaluated for one operating point is applicable for solving the OPF problems for many other operating points, with the condition that there are no drastic changes in the line flows \cite{guointelligent}. 

%As the definition of the affinity matrix in (\ref{eq:affinity}) is based on the optimality conditions of the considered problem and the physical topology of the system, it can be obtained for different problems on different systems. 
As most distributed methods solve for the optimality conditions in either a direct (such as OCD) or indirect (such as ADMM) way, one can follow similar analysis on the optimality conditions to define an affinity metric that works effectively with a specific distributed algorithm. In fact, we have used the same affinity matrix in \cite{guointelligent} in combination with OCD but there specifically derived the matrix based on the convergence criterion of OCD. The derivation above provides a more general justification for the usefulness of this affinity matrix also for other distributed approaches which is important for the generalization of the partitioning technique. 

Note that here we only provide one definition of the affinity matrix that works well empirically, which, however, is not the only way to define the affinity matrix. In fact, for the considered OPF problem, constructing the affinity matrix by using only the admittance matrix $Y$ can also yield good partitions due to the fact that the entries in $Y$ are close to many entries in $H$. How to construct the optimal affinity matrix is an open question in spectral clustering, and is a subject of future studies in the context of power system partitioning. 

%Another reason that using $H$ in the construction of the affinity matrix is effective for ADMM is the following: In ADMM, an agreement on the voltages at the boundary buses needs to be reached. If the resulting Lagrange multipliers associated with the power balances at the boundary buses are large, the mismatch between the duplicate voltages determined by different regions would have more significant impact on the objective function. Fortunately, $H$ contains the information on Lagrange multipliers associated with bus power balances. Using the defined affinity matrix, the buses with large Lagrange multipliers are less likely to appear at the boundaries in the resulting partition.

%Note that here we only provide one definition of the affinity matrix that works reasonably well, however, better partitions might be devised by refining the affinity matrix according to the properties of the chosen distributed algorithm.
\begin{algorithm}[t]
\caption{Spectral Partitioning}
\label{algIP}
\begin{algorithmic}[1]
\State \textbf{Input} System configuration and number of regions $K$
\State \textbf{Output} The assignment of each bus to the $K$ regions
\State Derive the affinity matrix for the entire system
\For{k=1 to $N$}
\State Perform spectral clustering based on the affinity matrix \indent to cluster the buses into $K$ regions with the K-means \indent method as the last step of spectral clustering
\EndFor
\State Choose the most balanced partition from the $N$ solutions
\end{algorithmic}
\end{algorithm}

%% file: ADMMResult.tex
\section{Simulation Results and Discussion}
\label{ADMMresult}
In this section, three sets of simulation results are presented. First, the partitions computed by the spectral partitioning method are compared with other partitions used in the literature in terms of the resulting performance of ADMM. Then, the impact of the number of regions on ADMM is presented. Finally, the performance of ADMM is evaluated after line flow limits are added to the OPF problem.
\subsection{Simulation Setup}
The test case used is the Polish 2383-bus system during the winter period, which contains 327 generators and 2896 transmission lines,  and its configuration is taken from MATPOWER \cite{zimmerman2011matpower}. The SNOPT package in TOMLAB \cite{Holmstrom97tomlab--}, which is an efficient solver for large-scale non-convex optimization problems, is used for solving the local OPF subproblems at each ADMM iteration. The simulations are run on a MacBook Air in MATLAB v8.5.

The algorithm is started from two different initial conditions, and we refer to these cases as the warm start case and the flat start case. In the warm start case, the starting point is a feasible power flow solution provided by MATPOWER. The warm start is a common choice in practice to solve large-scale OPF problems. In the flat start case, the voltages at all buses are initially set to $1$ p.u., from which convergence is typically hard to achieve. For both cases, the initial Lagrange multiplier estimates are set to zero. The initial penalty parameter $\rho_0$ and its incremental step size $\tau$ are set to $10^7$ and $1.1$, respectively, in the warm start case to achieve fast convergence, while in the flat start case, they are set to lower values. More details about this and about the complications associated with starting ADMM from a remote point are provided in Section \ref{flatstart}. Parameters $\gamma$, $\beta^{+}$, and $\beta^{-}$ are set to 0.9, 0.5, and 2, respectively. Convergence is declared when the largest primal residue $\|A_{k}{x}_{k}-{z}_{k}\|_{\infty}$ falls below $10^{-4}$ and the largest bus power mismatch falls below 0.01 MVA. 

\subsection{Evaluation of Partitions}
In this section, the impact of different partitions on the performance of ADMM is evaluated. The partition obtained using spectral partitioning is compared with the partition used in \cite{erseghe2015distributed}, which is based on the electrical distance between buses. Specifically, in \cite{erseghe2015distributed}, the partition is chosen in a way such that each load bus is assigned to the same region as its closest generator bus in terms of line impedance. If a total of $K$ regions is desired, $K$ generator buses are selected uniformly as the centers of each region. The distance between any two buses is then defined as the length of the shortest path in terms of line impedances. This partition considers the topology and electrical properties of the system, and hence it is a good benchmark for evaluating the performance of the spectral partitioning technique. This partitioning method based on electrical distance is denoted as EP in the following analysis, while the spectral partitioning method is denoted as SP. 

\subsubsection{The Warm Start Case}
\begin{figure}[t]
\setlength{\abovecaptionskip}{0.2cm} 
\centering
\captionsetup[subfigure]{captionskip= 0 cm}
\subfloat[40 regions, constraint violations]
{\label{fig:40ADMM}
\includegraphics[trim = 0mm 0mm 15mm 0mm, clip=true,width=4.0cm]{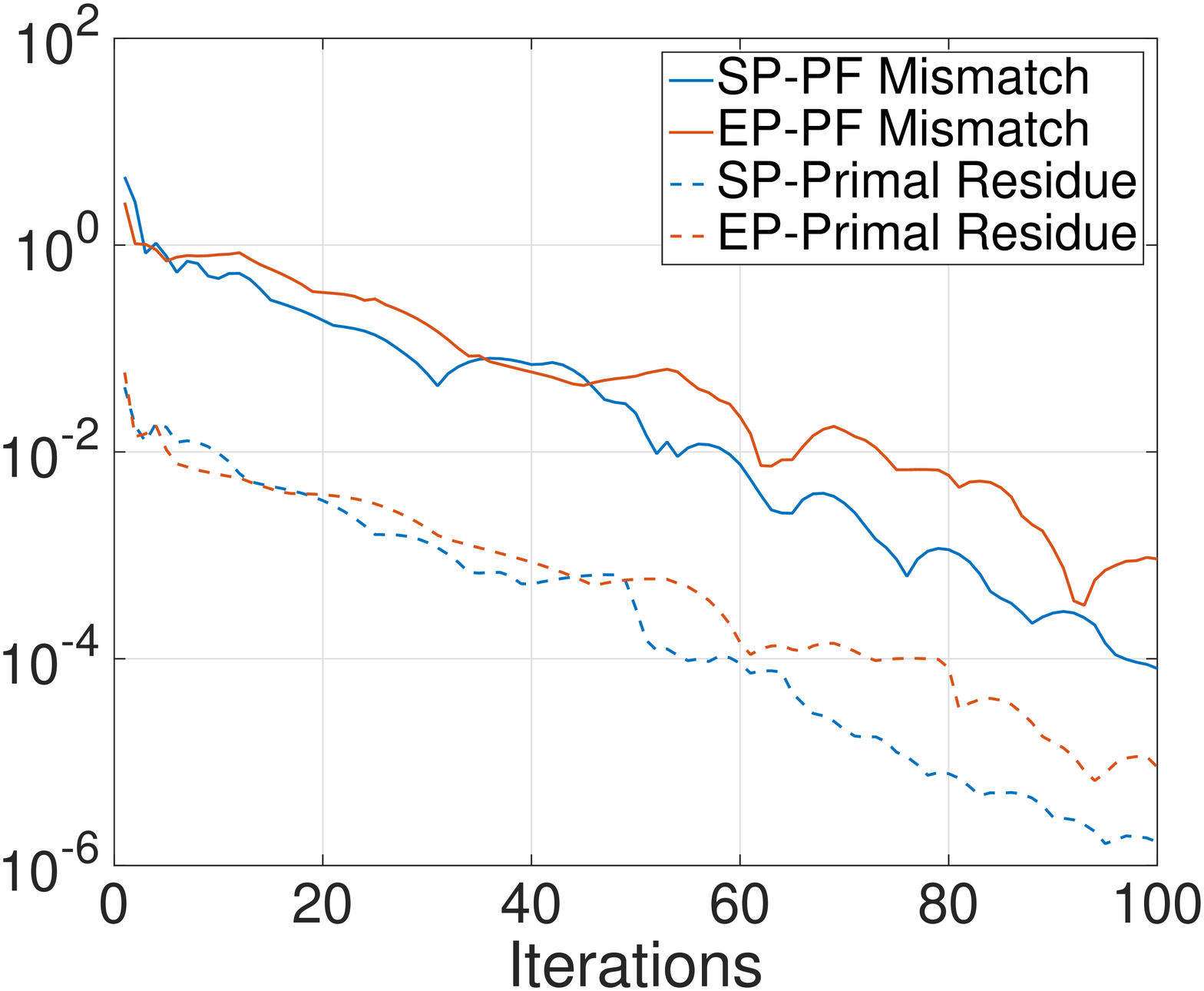} 
}
\hspace{0cm}
\subfloat[90 regions, constraint violations]
{
\label{fig:90ADMM}
\includegraphics[trim = 0mm 0mm 15mm 0mm, clip=true,width=4.0cm]{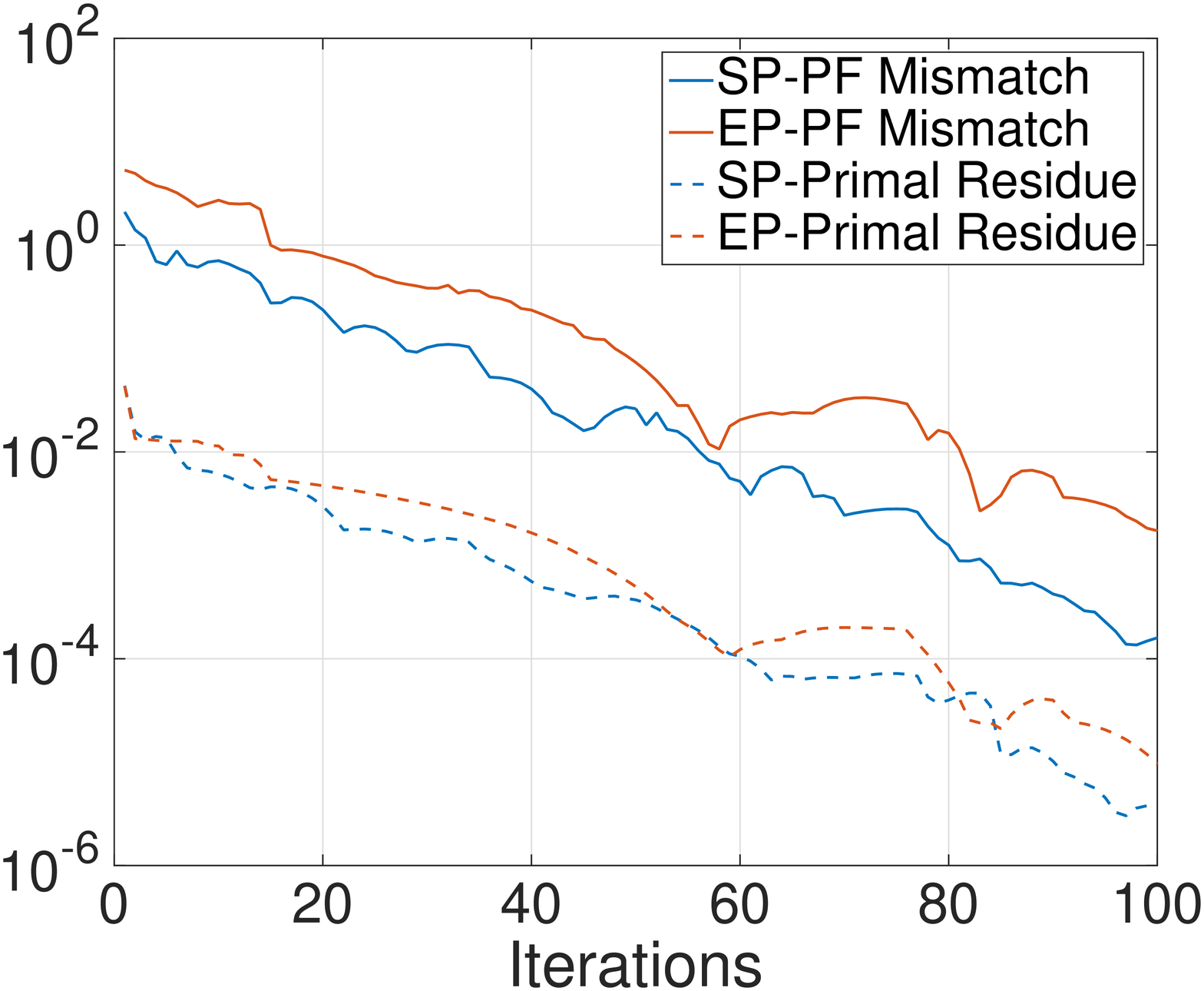} 
}\\
\subfloat[40 regions, gap in objective value]
{
\label{fig:40gapvsk}
\includegraphics[trim = 0mm 0mm 15mm 0mm, clip=true,width=4.0cm]{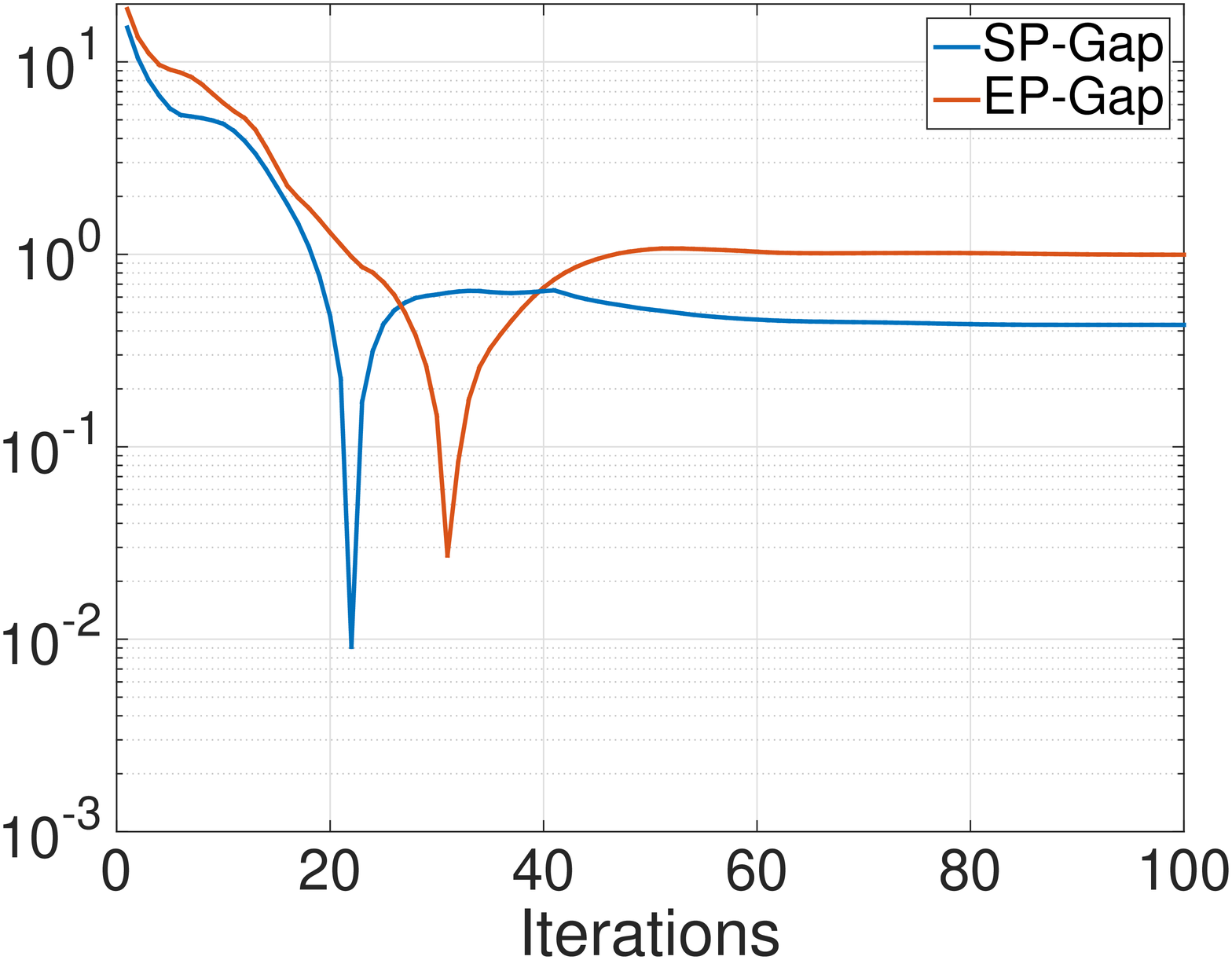} 
}
\hspace{0cm}
\subfloat[90 regions, gap in objective value]
{
\label{fig:90gapvsk}
\includegraphics[trim = 0mm 0mm 15mm 0mm, clip=true,width=4.0cm]{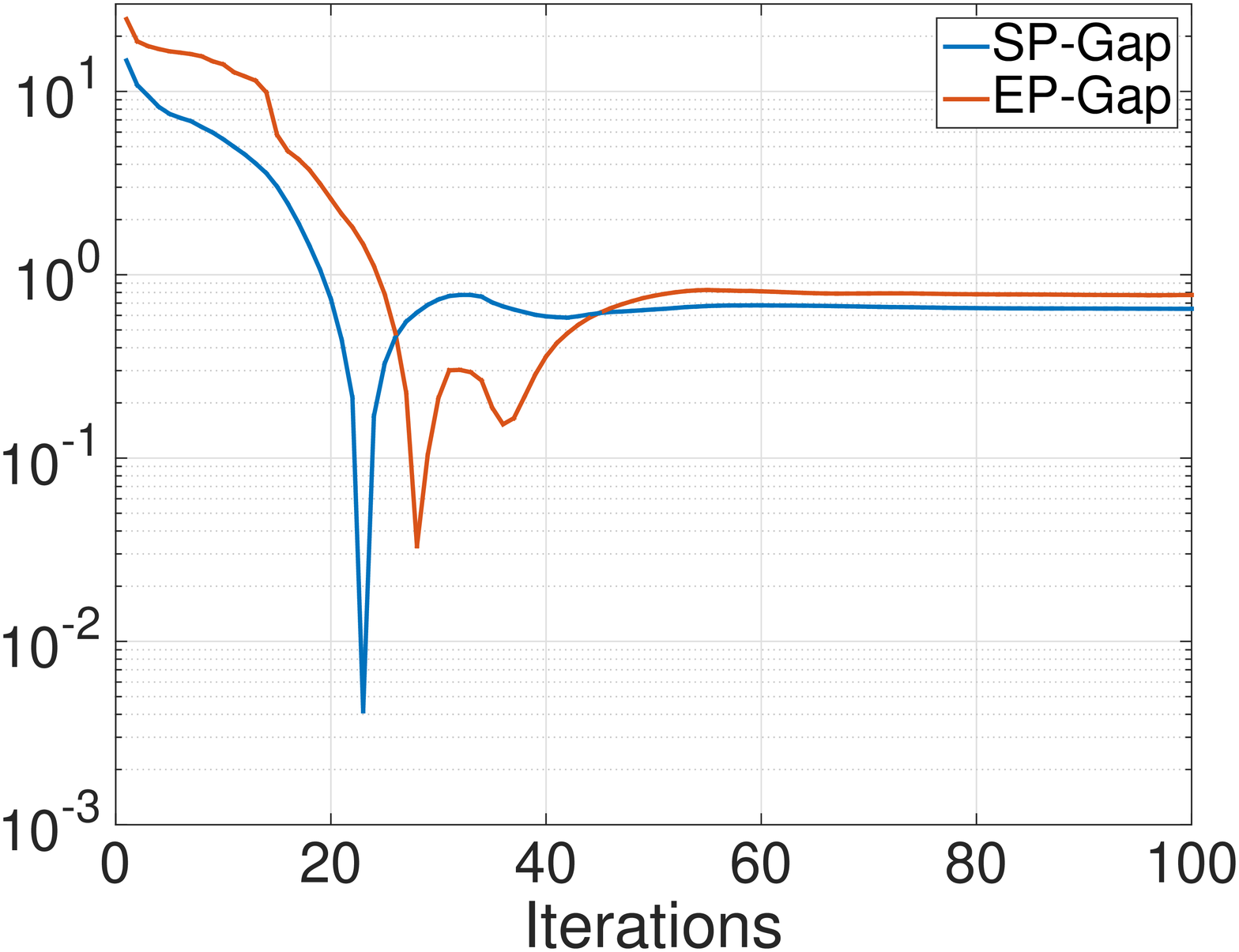} 
}
\caption{Convergence of ADMM with different partitions.}
\label{fig:ADMMConvergeOPvsSP}
\end{figure}

Figure \ref{fig:ADMMConvergeOPvsSP} shows the performance of ADMM with the partitions obtained by SP and EP, respectively, using a warm starting point. The 'Iterations' on the x-axis denotes the ADMM iteration as defined by (\ref{eq:ADMMIter}). The Polish system is divided into 40 and 90 regions using both methods. In Fig. \ref{fig:40ADMM} and \ref{fig:90ADMM}, the blue curves show the progress of ADMM with the partition obtained by SP, while the red curves show the progress of ADMM using EP partitions. The solid curve shows the largest bus power mismatch which is the maximum power flow balance violation, and the dotted curve shows the largest primal residue. With both 40-region and 90-region partitions, the blue curves are generally lower than the red curves, which shows that ADMM with SP approaches feasibility faster than with EP. Figures \ref{fig:40gapvsk} and \ref{fig:90gapvsk} show the relative error (in $\%$) of the objective value achieved by the distributed method with respect to the one obtained by a centralized method, which is denoted as the gap in the objective value. For non-convex problems, ADMM is not guaranteed to converge to a local optimum, which explains the difference in achieved objective values shown in the figures. As shown in Fig. \ref{fig:40gapvsk} and \ref{fig:90gapvsk}, ADMM with SP converges faster than with EP and achieves a smaller gap. 

It is worth noticing in Fig. \ref{fig:40ADMM} and \ref{fig:90ADMM} that the primal residue reaches $10^{-4}$ first compared with the power flow mismatch. This is due to the fact that in ADMM, the primal residue represents the difference between the copies of the voltages at the same bus. Since voltage errors propagate to line flow errors through multiplication by line admittances, it is necessary to add power flow feasibility checks in the termination condition of ADMM.

\begin{table}[b]
\caption{Comparison of different partitions with ADMM.}
\vspace{-0.2cm}
\centering
\begin{tabular}{ p{1cm}<{\centering}| p{1.4cm}<{\centering}|p{1.4cm}<{\centering}| p{1.4cm}<{\centering}|p{1.4cm}<{\centering} }
\toprule
Partition&SP, 40 regions&EP, 40 regions&SP, 90 regions&EP, 90 regions\\
 \midrule
Iterations &97&119&110&128\\
\midrule
Time ($s$)&218&2442&133&1039\\
\midrule
Gap&0.43\%&1.10\%&0.65\%&0.78\%\\
\bottomrule
\end{tabular}
\label{ADMMpartitioncompare}
\end{table}

Table \ref{ADMMpartitioncompare} further compares the effects of partitions on the number of iterations, the estimated computation time, and achieved objective value of ADMM. Here, the computation time is estimated by assuming that all subproblems are solved in parallel. Hence, it assumes that the time it takes to solve all subproblems at each ADMM iteration equals the maximum time needed to solve a single one. Note that the time spent on information exchange is not accounted for in this paper, which will be investigated in future works. The number of iterations and computation time measure the efficiency of the distributed method, while the gap measures the quality of the solution achieved by the distributed method. As shown in Table \ref{ADMMpartitioncompare}, the computation time of ADMM using SP is significantly smaller than when using EP. The reason is that the regions defined by EP can be highly imbalanced, having several large regions that contain generators electrically close to many loads. As a consequence, solving the subproblems associated with those large regions can be time-consuming. In contrast, spectral partitioning aims to find balanced partitions \cite{ng2002spectral}, hence it usually generates more balanced regions. Furthermore, the solution quality achieved by ADMM is satisfactory as the gap stays below 1\% if SP is used.

\subsubsection{Robustness of Partitions}
\label{flatstart}
In this subsection, we evaluate the performance of ADMM with different partitions in the flat start case and with different penalty parameters. We expect that with a good partition, ADMM should also converge from a remote starting point and not be affected much by the choice of parameters.  
\begin{figure}[t]
\setlength{\abovecaptionskip}{0.2cm} 
\centering
\captionsetup[subfigure]{captionskip= 0 cm}
\subfloat[40 regions, the flat start case]
{
\label{fig:40ADMMwc}
\includegraphics[trim =5mm 0mm 15mm 0mm, clip=true,width=4cm]{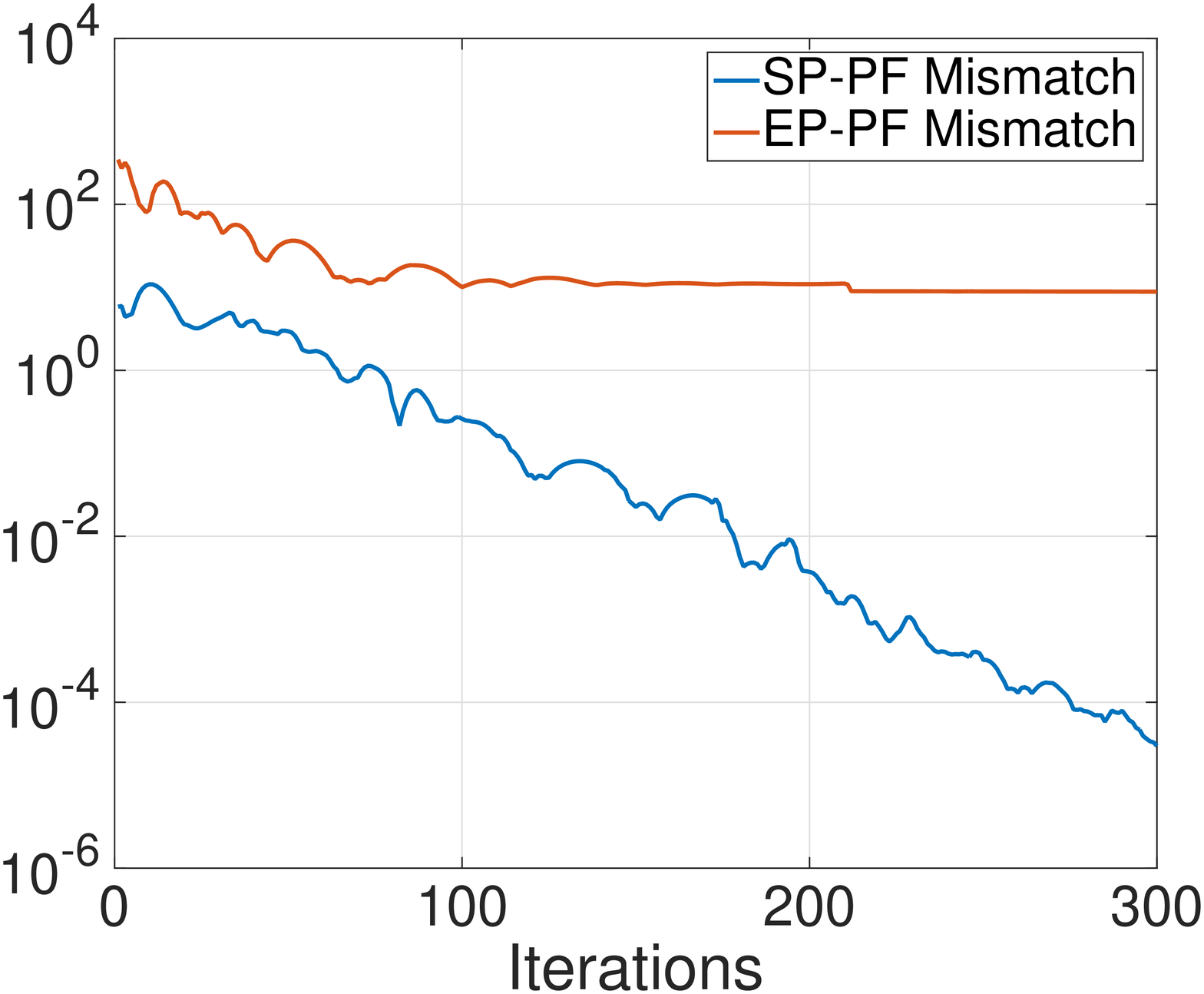} 
}
\hspace{0cm}
\subfloat[90 regions, $\rho_0=10^6$]
{
\label{fig:90ADMMsmalle}
\includegraphics[trim = 5mm 0mm 15mm 0mm, clip=true,width=4cm]{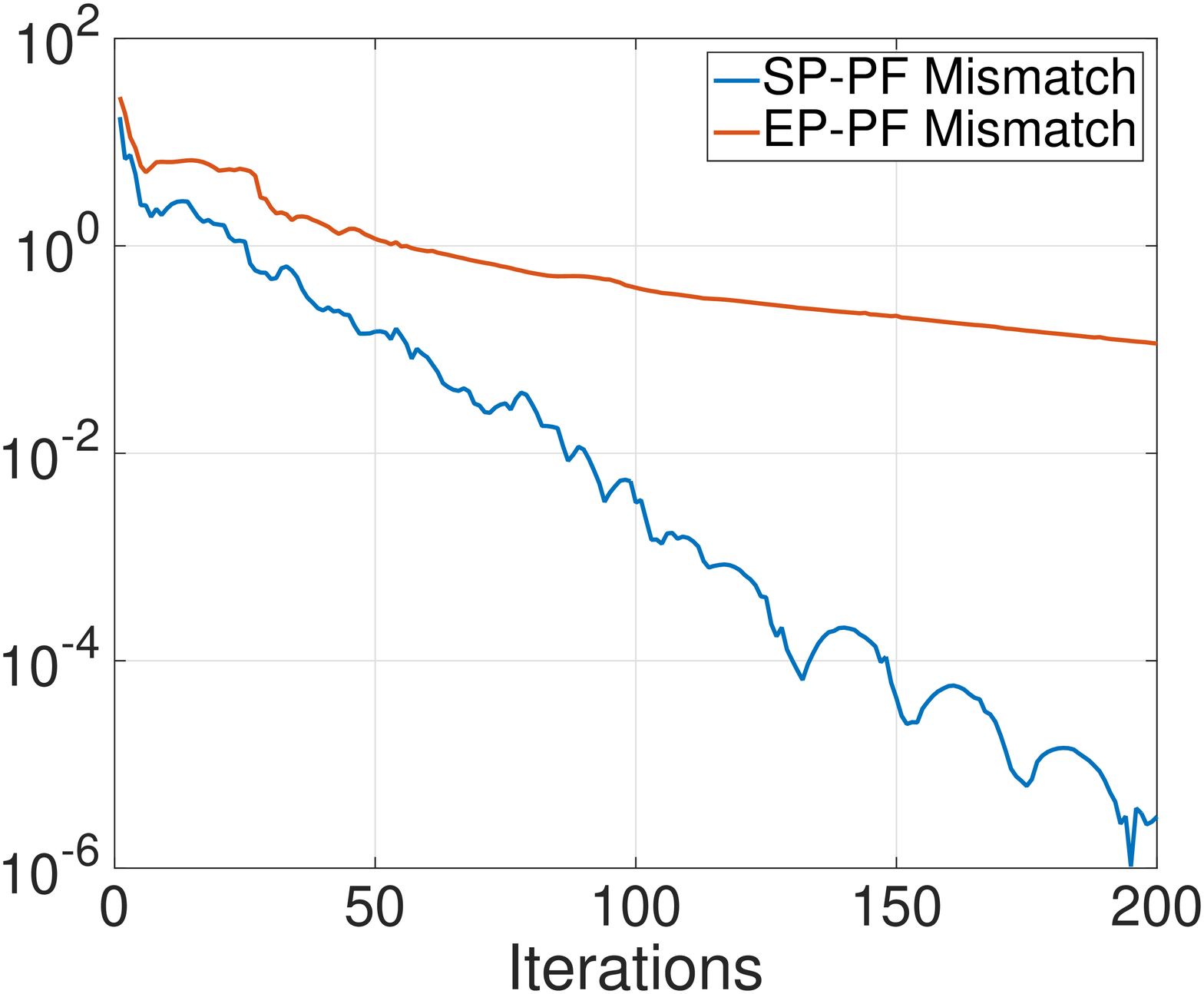} 
}
\caption{Mismatch in the power flow constraint with different partitions.}
\label{fig:RobustOPvsSP}
\end{figure}

In the flat start case, the initial penalty parameter $\rho_{0}$ and the incremental rate $\tau$ need to be set to lower values than those used in the warm start case. For the experiments, we use $\rho_{0}=10^{4}, 10^{5}$ and $\tau=1.05, 1.1$, respectively. A smaller value of $\rho_{0}$ may enable ADMM to converge if the algorithm runs long enough but will lead to significantly larger computation time which is impractical. The system is partitioned into 40 regions using the SP and EP methods. With the partition found by EP, ADMM fails to converge with all combinations of parameters stated above, while with the partition found by SP, ADMM only fails when $\rho_{0}=10^{5}$ and $\tau=1.1$. Figure \ref{fig:40ADMMwc} shows the best performance of ADMM out of the four settings of parameters using the two 40-region partitions computed by SP and EP, respectively, with a flat start. With SP, ADMM still converges to a feasible point, while with EP, ADMM fails to converge. For the solved case, ADMM takes 276 iterations, which results in 674 seconds, and the gap is 2.92 \%. Although the speed and accuracy of ADMM degrades in the flat start case compared with the warm start case, ADMM with the SP partition still converges to a feasible point. 

Figure \ref{fig:90ADMMsmalle} shows the performance of ADMM using different 90-region partitions with a different initial penalty $\rho_{0}=10^{6}$ in the warm start case. Again, ADMM only converges with the SP partition, while it fails to converge with the EP partition, which shows that ADMM is more robust with respect to the parameter settings using SP.

\subsection{Impact of the Number of Regions}
\label{impactregions}
\begin{figure}[t]
\setlength{\abovecaptionskip}{0.2cm} 
\centering
\captionsetup[subfigure]{captionskip= 0 cm}
\subfloat[Bus power mismatches]
{
\label{fig:curveall}
\includegraphics[trim = 0mm 0mm 15mm 0mm, clip=true,width=4cm]{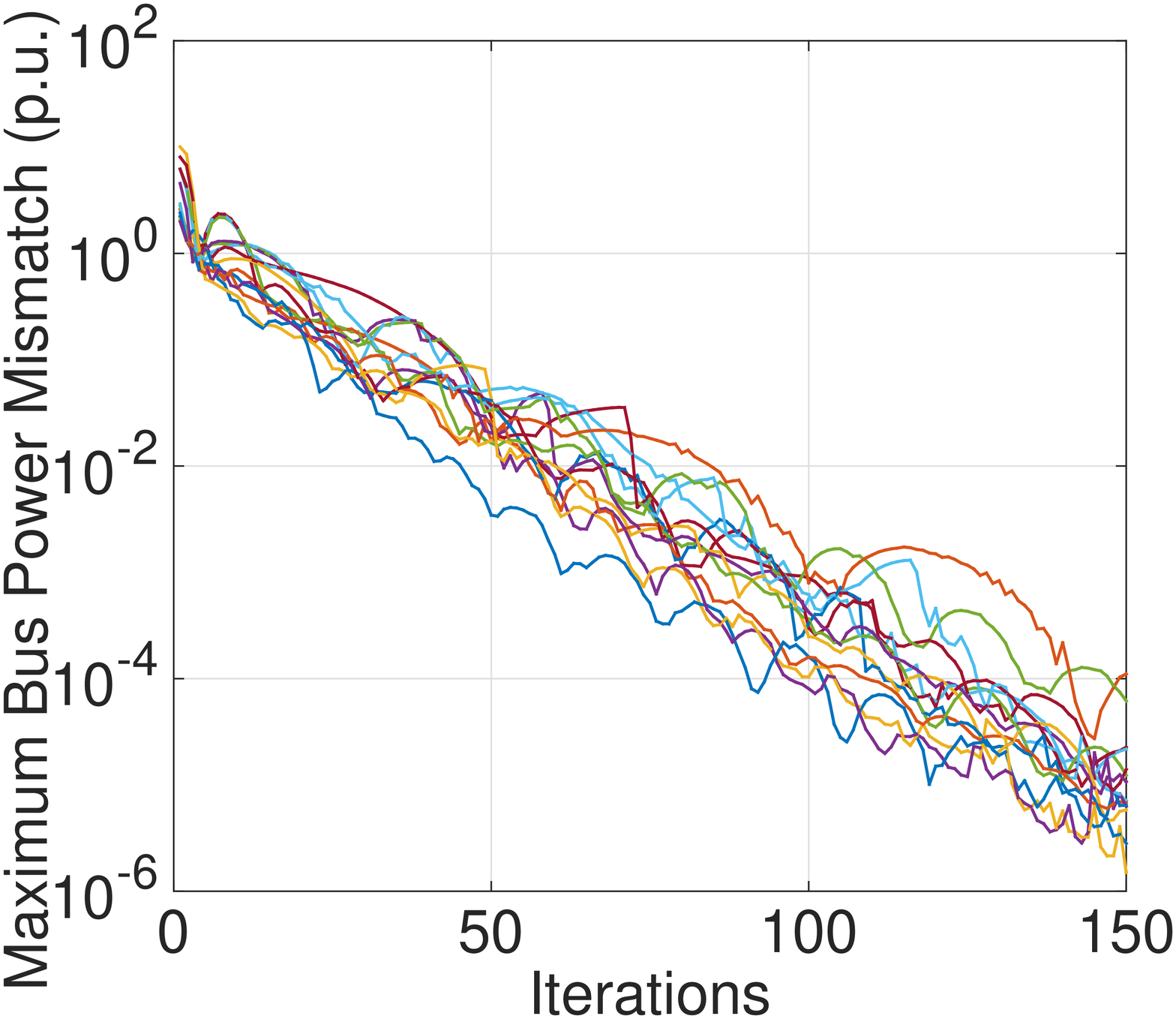} 
}
\hspace{0cm}
\subfloat[Number of iterations]
{
\label{fig:kall}
\includegraphics[trim = 0mm 0mm 15mm 0mm, clip=true,width=4cm]{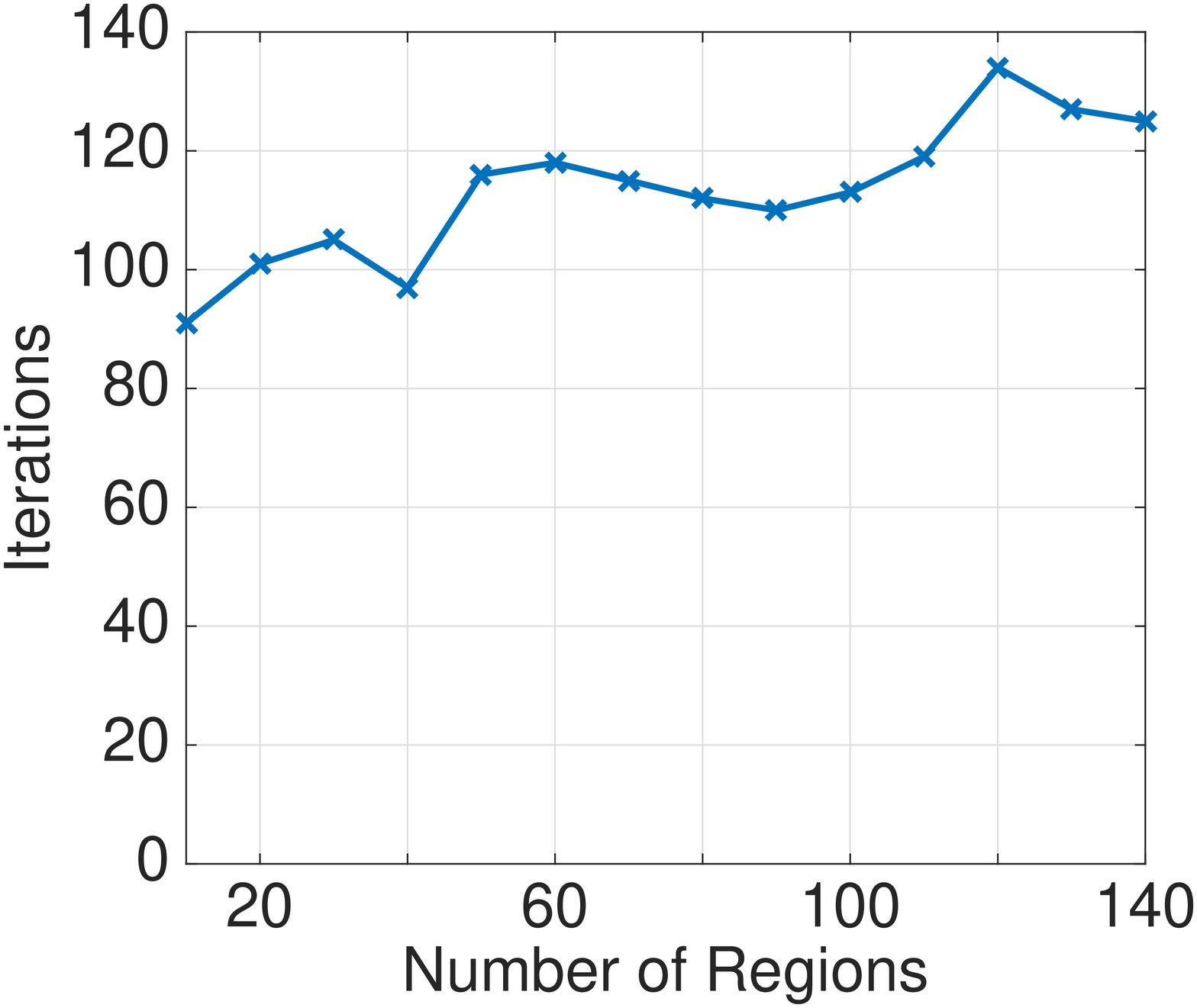} 
}\\
\subfloat[Computation time]
{
\label{fig:timeall}
\includegraphics[trim = 0mm 0mm 15mm 0mm, clip=true,width=4cm]{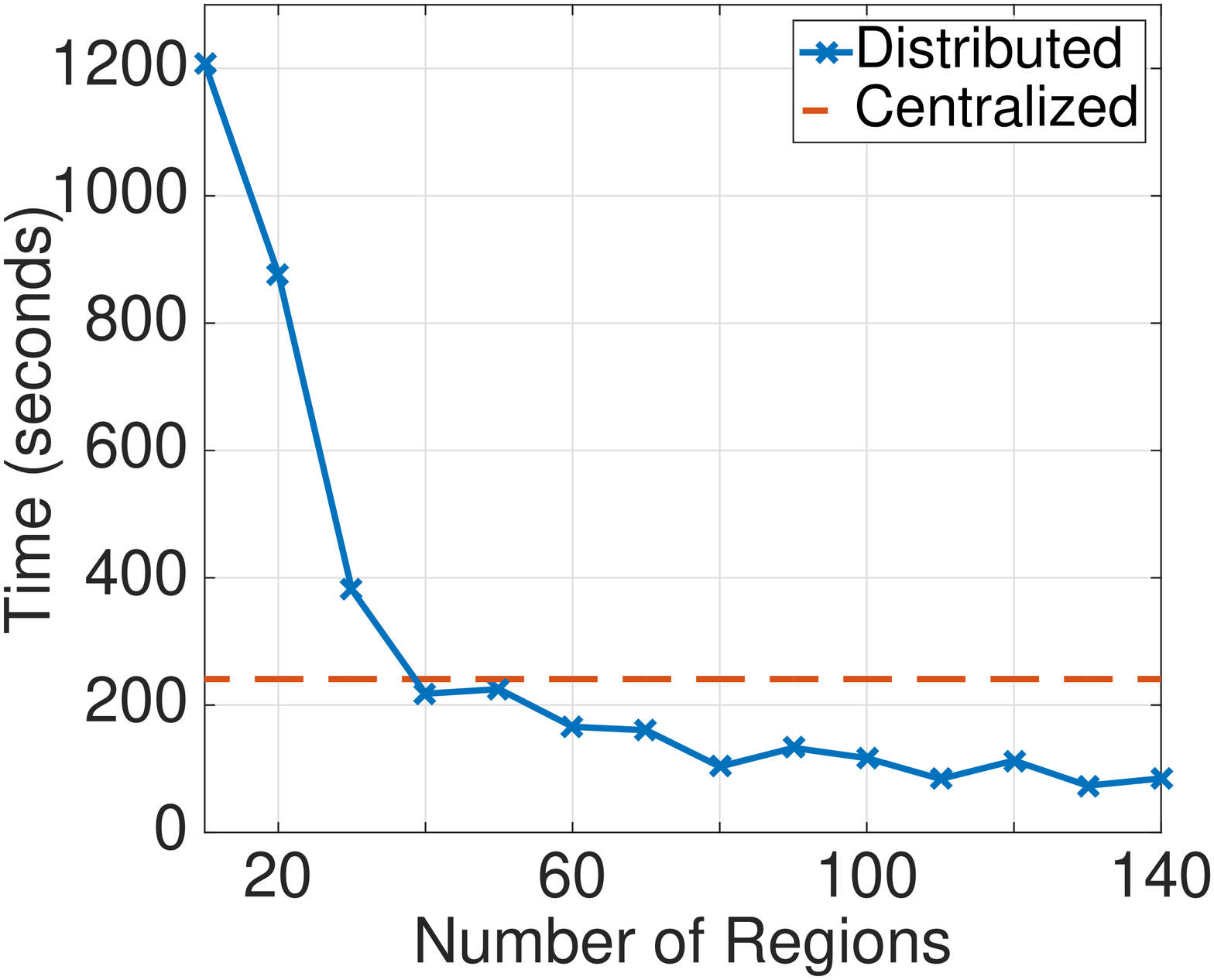} 
}
\hspace{0cm}
\subfloat[Gap]
{  
\label{fig:gapall}
\includegraphics[trim = 0mm 0mm 15mm 0mm, clip=true,width=4cm]{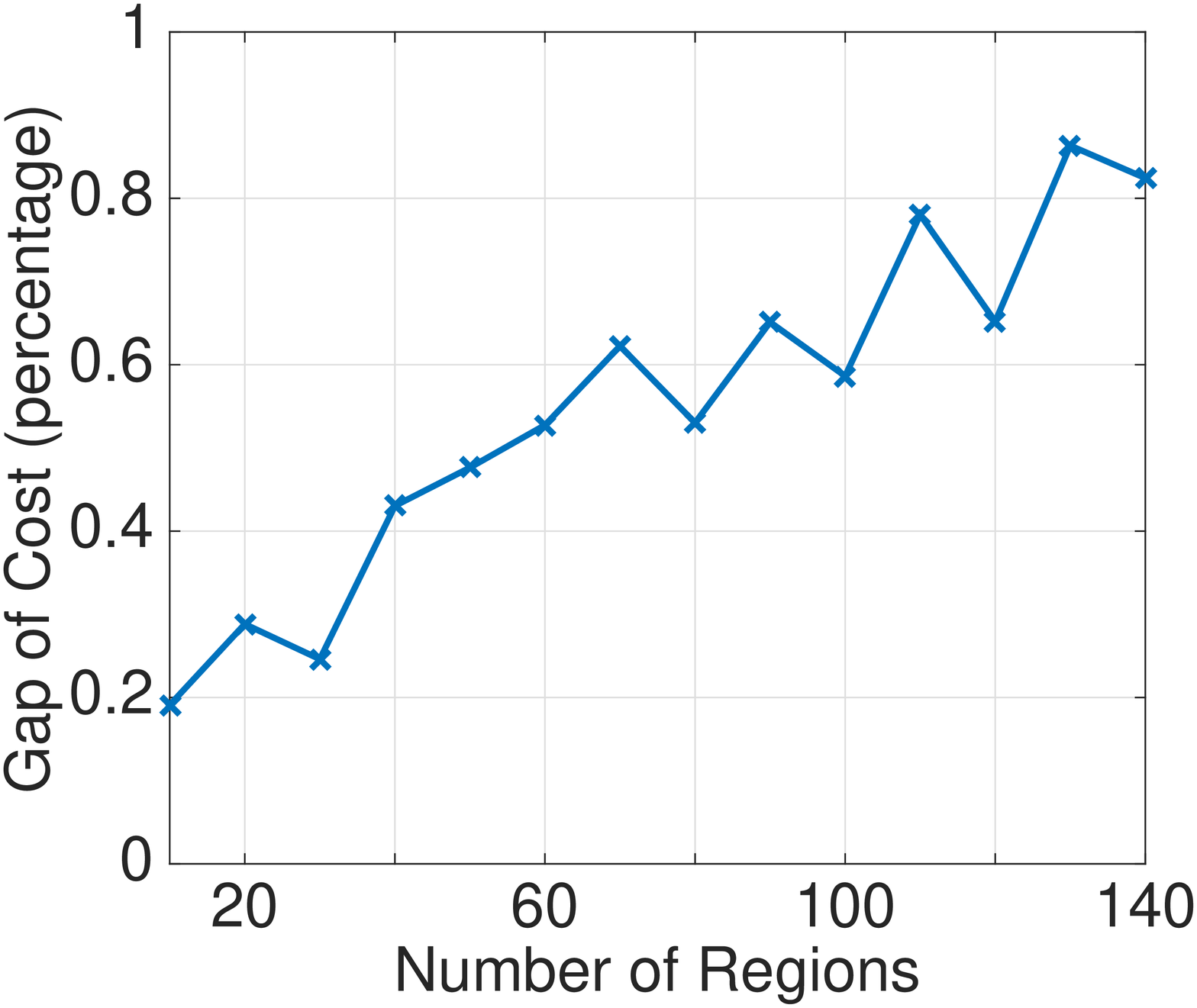} 
}
\caption{Performance of ADMM with different number of regions.}
\label{fig:ADMMall}
\end{figure}

In this section, we evaluate how the performance of ADMM changes as the number of regions changes. In the following experiment, the SP method is used to find partitions of the Polish system with 10 to 140 regions with an increment of 10 regions between any two partitions. These partitions cover both a mild partitioning of the system where there are hundreds of buses in a region and a severe partitioning of the system where there are only tens of buses in a region. The performance of ADMM with different number of regions is illustrated in Fig. \ref{fig:ADMMall} for the warm start case.

Figure \ref{fig:curveall} shows the decrease of the maximum bus power mismatch, where each curve corresponds to a partition with a different number of regions. Impressively, with all the partitions, a feasible point can be found within 150 iterations regardless of the level of partitioning of the system. Figure \ref{fig:kall} shows that the number of iterations does not increase significantly as the number of regions increases. Figure \ref{fig:timeall} shows the computation time with different number of regions, where the red dotted line denotes the time spent on solving the AC OPF problem in a centralized manner. As expected, the computation time of ADMM tends to decrease as the system is partitioned into more regions, since the time spent on solving subproblems can be greatly reduced if the scale of the subproblems is small. However, as shown in Fig. \ref{fig:gapall}, the gap increases with the increase of the number of regions. In Section \ref{discussion}, we will discuss more on how to determine the number of regions. 

\subsection{Adding Line Limits}
\label{ResultTL}
We now evaluate the performance of ADMM with thermal line limits included considering the following two scenarios. In the first scenario the line limits of all the transmission lines are added to the constraint set while in the second scenario only the line limits of several lines prone to congestion are added. With the partitions devised in Section \ref{impactregions}, there are tie lines congested in each of the partitions. To investigate whether the congestion of tie lines has significant impact on the performance of ADMM, another partition is devised by manually increasing the affinity between the buses that the congested lines connect to in the SP method. As a consequence, the obtained partition does not include any congested tie lines as each congested line is included in a single region.
\begin{table}[b]
\caption{Performance of ADMM with different partitions when all line limits are included.}
\centering
\begin{tabular}{ p{1cm}<{\centering}| p{2cm}<{\centering}|p{1.8cm}<{\centering}| p{1.8cm}<{\centering} }
\toprule
Partition&40 regions, no tie line congested&40 regions, tie line congested&80 regions, tie line congested\\
 \midrule
Iterations &161&124&147\\
\midrule
Time ($s$)&679&511&237\\
\midrule
Gap&1.92\%&1.56\%&2.25\%\\
\bottomrule
\end{tabular}
\label{ADMMallLL}
\end{table}
\begin{table}[hbtp]
\caption{Performance of ADMM with different partitions when few line limits are included.}
\centering
\begin{tabular}{ p{1cm}<{\centering}| p{2cm}<{\centering}|p{1.8cm}<{\centering}| p{1.8cm}<{\centering} }
\toprule
Partition&40 regions, no tie line congested&40 regions, tie line congested&80 regions, tie line congested\\
 \midrule
Iterations &129&125&127\\
\midrule
Time ($s$)&304&346&122\\
\midrule
Gap&1.05\%&0.96\%&1.25\%\\
\bottomrule
\end{tabular}
\label{ADMMfewLL}
\end{table}

Table \ref{ADMMallLL} shows the performance of ADMM with all line limits added. In this setting, the centralized approach takes 355 seconds to convergence, which is longer than the case without line limits. As shown in Table \ref{ADMMallLL}, with line limits ADMM requires more iterations and time to terminate and the optimality gap is also larger. This is due to the fact that solving the subproblems takes longer and an agreement on the voltages at the boundaries is harder to achieve with the additional line constraints. However, comparing the two 40-region partitions given in Table \ref{ADMMallLL}, we notice that the performances of ADMM are similar whether the congestion occurs on tie lines or non-tie lines. With the partition with tie lines congested, ADMM even performs better than with the partition where the congested lines are forced to be included in a single region. This observation suggests that the partition obtained by SP does not need to be changed even if congestion occurs on tie lines.

Better performance of ADMM can be achieved when only a few line limits are incorporated, as shown in Table \ref{ADMMfewLL}. The included line limits are the ones associated with the congested lines when all the line constraints are considered. With all the three partitions, a gap of around 1\% can be achieved. Hence, with only certain line constraints considered, ADMMM can still reach a high quality solution efficiently. 
\vspace{-0.2cm}
\section{Physical Interpretations of Results}
\label{discussion}
Our results provide additional evidence that the recently developed partitioning technique \cite{guointelligent} in conjunction with the ADMM method can be used for distributed optimization in real-world large-scale power systems. While the evidence shown in this study is not sufficient to guarantee that this approach would work effectively for all non-convex problems and for other decomposition techniques, it is a promising step in this direction. 

As demonstrated in Section \ref{ADMMresult}, the efficiency and robustness of a distributed method is highly dependent on the system partitioning method used. Fortunately, the spectral partitioning technique first proposed in \cite{guointelligent} and used in this study is indeed very effective in improving the efficiency and robustness of distributed methods.

% which could alleviate the difficulty in reaching a solution using distributed methods in large-scale power systems. Computation efficiency is considered as one of the main benefits of distributed methods, which can only be realized if the partitioning of the system is chosen properly. The robustness of distributed methods is also a valuable property in practice, as the starting points of the algorithm and operating points of the system might always change, and it is usually hard to tune the parameters of distributed algorithms. 

It can be observed from the results in Section \ref{ADMMresult} that the performance of the distributed method also depends on the number of regions. For example, in Fig. \ref{fig:timeall}, there is a cross-over point (around 40 regions) of the solid and dotted curves beyond which the computation time of the distributed approach is smaller than the centralized approach. This shows that in terms of the estimated computation time, there is a critical number of regions in implementing the distributed method to outperform the centralized approach. However, this does not necessarily imply that one should partition the system into as many regions as possible because more regions imply more communications, which is an important concern in distributed optimization that will be investigated in future works. While the time needed for information exchange is not accounted for in this paper, it is clear that partitioning the system into more regions would require more communications among the regions, which may lead to large delays. In addition, a partition with more regions would require more regional computational entities. Hence, the number of regions is a design parameter that should satisfy specific requirements of the utilities in terms of both the performance of the distributed method and the available computational and communication resources. 

While the results of this study look very promising, the limitation of the presented approach is that it can only find solutions close to \textit{local optimums} of non-convex problems, which, however, is also a general limitation of most distributed optimization approaches applied to non-convex problems. Further research is needed to address this limitation of the presented approach.

%While the results of this study look very promising, the limitation of the presented approach is that it can only find local optimums of non-convex problems, which, however, is also true with centralized approaches. Another limitation is that the distributed approach can only identify a solution close to local optimums with a gap in the objective function. Further research is needed to address these limitations in the presented approach.
Possible future work includes evaluation of the distributed optimization approach on large-scale systems with communications between different regions incorporated. In particular, it will be interesting to investigate how different communications configurations will affect the delay performance of the distributed approach. 
%It should be mentioned that the centralized approach solves the AC OPF problem in 241 seconds. Hence, only if the partitions determined by the IP method are used, the distributed approach is more efficient than the centralized approach assuming no communications delay.
%
%These results indicate that with the partitions of IP, the distributed method is less sensitive to the change in starting points or the algorithm parameters, which is a valuable property in practice, as the starting points might always change and it is usually hard to tune the parameters of the distributed algorithms.
%
%The Polish system is meshed and complex, and distributed methods are known to have difficulty in converging in such harsh situations \cite{erseghe2015distributed}. Hence, this further validates the contributions of the proposed partitioning method as it provides partitions with more balanced and decoupled regions, which alleviates the difficulty in reaching a solution in a distributed manner.

%% file: ADMMConclusion.tex
\section{Conclusions}
\label{ADMMconclusion}
In this paper, we applied a previously developed spectral partitioning method in conjunction with the distributed optimization method, ADMM, to solve a non-convex OPF problem of the Polish 2383-bus system in a distributed fashion. By using the partitions computed by the partitioning method, ADMM can find a solution very close to the local optimum efficiently, which shows the applicability of distributed optimization in large-scale real power systems. Additionally, the partitions found are robust with respect to the starting point, the parameters used in ADMM, and the inclusion of thermal line constraints. These results show that the partitioning method can be generalized to different distributed methods, e.g., ADMM, which provides a structured approach to decompose a large-scale system such that an efficient implementation of distributed optimization methods is possible. This is of high importance as it is a key enabler to make distributed methods practically viable.